\documentclass[a4paper,10pt]{article}

\usepackage{graphicx,soul}
\usepackage{amsmath}
\usepackage{epsfig}
\usepackage[usenames, dvipsnames]{color}
\usepackage{array}
\usepackage{authblk}

\title{Density-based clustering: A `landscape view' of multi-channel neural data for inference and dynamic complexity analysis}
\author[1]{Gabriel Baglietto}
\author[2]{Guido Gigante}
\author[3]{Paolo Del Giudice}

\affil[1]{\small INFN-Roma1, Rome, Italy and IFLYSIB, Instituto de F\'isica de  L\'iquidos y Sistemas Biol\'ogicos (UNLP-CONICET), La Plata,  Argentina}
\affil[2]{\small Italian Institute of Health, Rome, Italy and Mperience srl, Rome, Italy}
\affil[3]{\small Italian Institute of Health and INFN-Roma1, Rome, Italy}

\begin{document}

\maketitle

\begin{abstract}
 
Simultaneous recordings from N electrodes generate N-dimensional time series that call for efficient representations to expose relevant aspects of the underlying dynamics.

Binning the time series defines a sequence of neural activity vectors that populate the N-dimensional space as a density distribution, especially informative when the neural dynamics proceeds as a noisy path through metastable states (often a case of interest in neuroscience); this makes clustering in the N-dimensional space a natural choice. 

We apply a variant of the `mean-shift' algorithm to perform such clustering, and validate it on an Hopfield network in the glassy phase, in which metastable states are largely uncorrelated from memory attractors. 

The neural states identified as clusters' centroids are then used to define a parsimonious parametrization of the synaptic matrix, which allows a significant improvement in inferring the synaptic couplings from the neural activities.
 
We next consider the more realistic case of a multi-modular spiking network, with spike-frequency adaptation inducing history-dependent effects; we develop a procedure, inspired by Boltzmann learning but extending its domain of application, to learn inter-module synaptic couplings so that the spiking network reproduces a prescribed pattern of spatial correlations.

After clustering the activity generated by such multi-modular spiking networks, we cast their multi-dimensional dynamics in the form of the symbolic sequence of the clusters' centroids; this representation naturally lends itself to complexity estimates that provide compact information on memory effects like those induced by spike-frequency adaptation. Specifically, to obtain a relative complexity measure we compare the Lempel-Ziv complexity of the actual centroid  sequence to the one of Markov processes sharing the same transition probabilities between centroids; as an illustration, we show that the dependence of such relative complexity on the characteristic time scale of spike-frequency adaptation.
\end{abstract}
\newpage

\section{Introduction}
\label{sec:intro}

Technology nowadays allows neuroscientists to simultaneously record brain activity from increasingly many channels, at multiple scales; indeed, recent years witnessed a kind of `Moore's Law' for neural recordings \cite{stevenson2011}, and this poses new challenges and opens new opportunities. 

One obvious challenge is to devise data representations that easily convey in a compact form the spatio-temporal structure of the recorded data.
Various forms of dimensional reduction are now commonly used in the analysis of multiple recordings. In general terms, if one views recorded experimental data as a matrix whose columns are the `feature vectors' (in the case at hand, the set of recorded activities in a given time bin), and whose rows span the `sample space' (in the case at hand, the successive time bins), dimensional reduction across the column direction provides a reduced representation in terms of few suitably identified features obtained from the original ones (e.g. principal component analysis); on the other hand, one can view clustering across the rows direction as a way to reduce the dimensionality of the data matrix by lumping together, according to some similarity measure, groups of activity vectors sampled at different times. This latter viewpoint, which we take here, to our knowledge has been much less used in neuroscience.

On the other hand, one recently explored opportunity is to take advantage of multiple recordings to revive old approaches to infer estimates of synaptic couplings from measured correlations between neural activities. Correlations measured from single neuron pairs obviously can only provide ambiguous estimates of the direct synaptic couplings (due to confounding causes like common input to the sampled neurons). However it was noticed in a landmark paper \cite{schneidman2006weak} that when many (order 100 for instance) simultaneous recordings are available, even though the underlying biological neural network is still dramatically undersampled, the global pattern of (individually small) pairwise correlations allows to extract meaningful information about the synaptic connectivity. This was achieved by assuming a maximum entropy Ising model, for which an ``inverse Ising'' problem was solved to infer the parameters (couplings and external input) for the given data. Many efforts were subsequently devoted both to extend the approach to non-equilibrium estimates, and to lighten the computational load of maximum entropy estimates (Boltzmann learning) through various mean-field approximations (see {\it e.g.} \cite{hertz2013}\cite{roudi2015multi}\cite{capone2015inferring}).

In the present work we propose an approach, based on clustering in the multi-dimensional state space of simultaneous recordings, that provides both an advantage for a compact representation of data, also amenable to efficient estimation of the complexity of the system's dynamics, and besides allows to improve inference on the network couplings.

After describing the clustering method (which is a slightly modified version of the `mean-shift' algorithm \cite{rodriguez2014clustering}\cite{Fukunaga}), we first illustrate its working on time series generated from the dynamics of a Hopfield network which, since it possesses an energy function, naturally lends itself to density-based clustering in the state space; here we choose a `hard' regime where the Hopfield network is in a disordered phase and spatio-temporal structures related to the energy landscape are not easily discernible from the time series. At this stage we also formulate a parametrization of the model's synaptic matrix, based on the identified clusters, and show that it allows to obtain an inference of the synaptic couplings which is much more insensitive to noise.

We then move on to the more complex and biologically motivated case of a multi-modular attractor spiking network: its integrate-and-fire neurons are endowed with spike-frequency adaptation (SFA), acting as an activity-dependent self-inhibition that introduces history-dependent effects making the  `landscape' dynamic; modules are approximately bistable between high and low activity states; within-modules connectivity is stronger than between-modules.
 
Reasons to choose this particular context include recently published evidence \cite{MattiaEtAlJNeurosci2013}\cite{latimer2015single} that cortical dynamics can occur in the form of abrupt switches, between an `Up' and a `Down' states, of local self-excited modules, such that the whole dynamics appears as the evolution of a `binary word', each bit being the `Up' or  `Down' state of each cortical module.

In the chosen setting of weakly coupled bistable modules, the network dynamics is largely determined by inter-modular couplings; in choosing the latter we wanted to preserve some a priori knowledge of key features of the state space, to be later checked against the found cluster structure, and we do this by seeking inter-modular couplings such that the network possesses a prescribed pattern of spatial pairwise correlations, derived in turn from a template Hopfield network. 

To achieve this we develop a procedure, inspired by Boltzmann learning but extending its domain of application, which we believe has an interest beyond the contingent purpose.

We check the good performance of this new learning procedure by direct comparison between the prescribed spatial correlations and those generated by the optimal network at the end of learning; by comparing the state-space cluster structure of the optimal network and the template Hopfield network; by inferring effective inter-module synaptic couplings from simulations of the optimal network and comparing them to the synaptic efficacies of the template model.

Knowledge of the cluster structure also allows to cast the multi-dimensional time series generated by the network dynamics in the reduced form of a symbolic dynamics in the clusters' centroid space. We perform such a reduction to show that it allows to expose in a compact way time-dependent features like history-dependent effects; we do this by exploring a large range of characteristic times of SFA, and showing that the Lempel-Ziv complexity measure, applied to the centroid sequence, nicely captures the memory effects induced by SFA.

In summary, what we propose here is a way to use knowledge of the spatial correlations to develop informative and compact representations of multi-dimensional neural data, also allowing for improved inference and useful reduced representation of the multidimensional dynamics, and to develop a strategy for data-driven model building with spiking networks.

\section{Results}
\label{sec:results}

\subsection{When the landscape is (partially) known: The Hopfield Model as a first test ground for state-space clustering}
\label{sec:results-hopfield}

\begin{figure}[h!]
\setlength{\unitlength}{\textwidth}
\begin{picture}(1.0,0.6)
\put(0.0,0.0)
  {
\epsfig
   {
   file=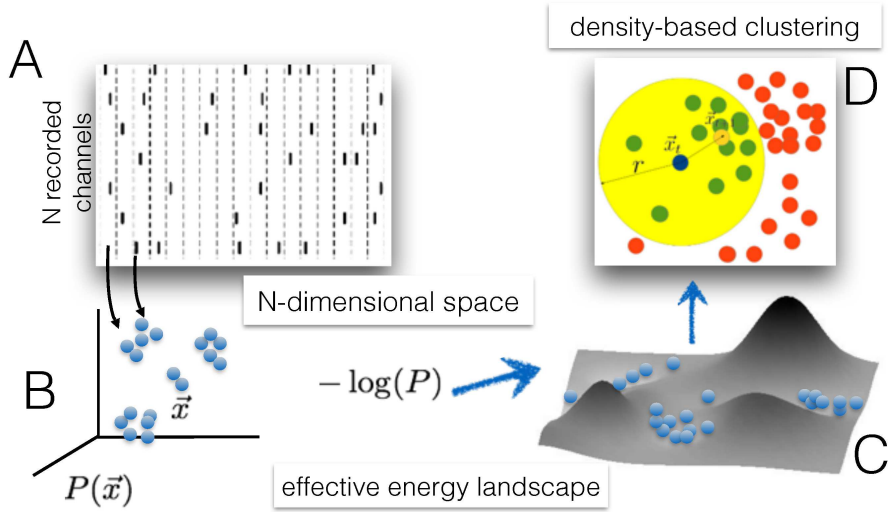,
   width=1\unitlength,
   }
  }
\end{picture}
\caption{Sketch of the  steps involved in the density-based clustering of multi-dimensional time series through the mean shift algorithm. The time-binned $N$-dimensional data (`spikes') in panel A define a density profile in a $N$-dimensional space, interpreted as sampling a stationary probability distribution (panel B), from which an effective energy landscape can be defined (panel C). Panel D illustrates how the mean-shift operates to find the local maxima of the density distribution. Blue dot: current position of a point to be moved by the algorithm. Big yellow area: the chosen neighbourhood of the blue point.  Green dots: neighbouring points of the blue point. Red dots: points that are not neighbours of the blue point. Small yellow dot: the new position of the blue point is given by the algorithm as the center of mass of its neighbouring points. See Section \ref{sec:methods} for details.}
\label{figure1}
\end{figure}

We start by providing an example of the information gained from clustering in the state space, {\it i.e.} finding the local maxima of the density distribution of the configurations generated by a system's dynamics, or - equivalently - the minima of an effective energy landscape (see Fig.~\ref{figure1}).

Though we want to ultimately apply clustering to data from realistic neural simulations, to illustrate the method we consider here a well known neural network (the Hopfield model) for which an equilibrium distribution of states is defined.  

For the Hopfield model \cite{Hopfield}, the equilibrium probability distribution of the neural states $\boldsymbol{\sigma} = \{\sigma_1, \, \sigma_2, \dots \sigma_N\}$ ($\sigma_i = \pm 1$) is given by:
\begin{equation}
\label{eq3}
p(\boldsymbol{\sigma}) \propto \exp\Big( -\beta \, H[\boldsymbol{\sigma}]\Big) ,
\end{equation}
where $\beta{}$ is interpreted as the inverse of a temperature and $H$ is the energy function:
\begin{equation}
H[\boldsymbol{\sigma}]=-\frac{1}{N} \, \sum_{i,j = 1}^N \sum_{\mu=1}^P \xi_i^{\mu} \, \xi_j^{\mu} \, \sigma_i \sigma_j,
 \label{eq4}
\end{equation}
where the $P$ discrete vectors $\boldsymbol{\xi}^\mu{}$ are chosen as random uncorrelated configurations of the $N$ spins ($\xi^\mu_i = \pm 1$) which, depending on $P$, $N$ and $\beta$ can act as `stored patterns', i.e. the dynamics relaxes to the neighborhood of one of the configurations $\boldsymbol{\xi}^\mu{}$, which are minima of the energy (maxima of the probability distribution) together with their mirror patterns $-\boldsymbol{\xi}^\mu{}$ \cite{amit1992modeling}. Depending on $P$, $N$ and $\beta$, besides the patterns $\boldsymbol{\xi}^\mu{}$ a large number of energy local minima is also present, notably various linear combinations of the stored patterns themselves or states uncorrelated with the patterns in the `spin glass' phase.

In the following we consider the same Hopfield model (same synaptic couplings, therefore same energy minima) in two operating regimes: low noise (\textit{i.e.} temperature, where the network is expected to reside most of the time in the proximity of the deepest energy minima), and high noise (where the network explores larger regions of the state space, less constrained by the structure of the underlying energy function). The network comprises $N=50$ neurons and $P=4$ stored patterns. Although for such small numbers relying quantitatively on known theoretical estimates of memory capacity is unwarranted, for the chosen values of $N$ and $P$ the model would be predicted to be well above the retrieval phase for both temperatures, and for its non-trivial (`glassy') energy landscape the energy minima are not expected to coincide with the memory patterns. 

The neural configuration time series were generated simulating the Hofpield model for 20000 Monte Carlo steps (a suitable number of initial thermalization steps were neglected).

To the $N$-dimensional resulting time series we applied the (slightly modified) mean-shift (MS) clustering procedure as described in Section \ref{sec:methods}: essentially, we iteratively move the points in the $N$-dimensional space, each of which represents one state generated in the Monte Carlo sequence, towards the center of mass of their neighboring points, thereby identifying at the end the estimated position of the local maxima of the density distribution in the state space.

The states of the Hopfield model are defined over the $N$-dimensional hypercube, where $N$ is the number of neurons. In order to perform the mean shift displacements we take the sign of the mean value of each coordinate, so that the points remain in the original space. Any time the mean value of a given coordinate is zero, we don't shift that coordinate. 	

Finally, after reaching a convergence criterion (see Section \ref{sec:methods}) of the MS algorithm, we re-run it over the set of found centroids, with a fixed radius given by Hamming distance 2 (overlap 0.92) \footnote{The overlap $q^{\alpha \beta}$ between two binary configurations, and weighting each centroid with its mass in the analogous of Eq.~\ref{eq.MeanShift}.
$\sigma^{\alpha}$ and $\sigma^{\beta}$ is defined as 
$q^{\alpha \beta}=\frac{1}{N} \sum_{i=1}^N \sigma_i^{\alpha} \sigma_i^{\beta}$, and it is related with the Hamming distance $h^{\alpha \beta}$ by $q^{\alpha \beta}=1-2 \frac{h^{\alpha \beta}}{N}$.}; we found this further step to make clustering more resistant to noise.

In general, we will consider only clusters containing data points above a minimal fraction (``cutoff'') of the whole time series, \textit{i.e.} minimal {\it mass} (1$\%$ unless specified). In this way we avoid to consider as clusters small bumps of density due to finite sample fluctuations.

\begin{figure}[!h]
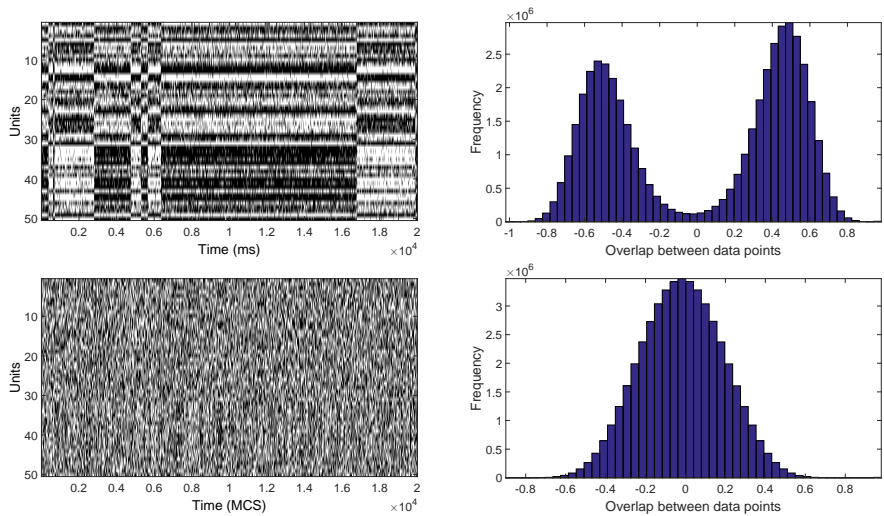

\begin{center}
\setlength{\unitlength}{\textwidth}
\begin{picture}(1,0.55)

\put(0.0,0.28)
  {
   \epsfig
   {
   file=fig2_Upper_Left_cold_rasterPlot.eps,
   width=0.45\unitlength,
   }
  }
   \put(0.5,0.28)
  {  
   \epsfig
   {
   file=fig2_Upper_Right_cold_Histogram.eps,
   width=0.45\unitlength,
   }
  }
  
  \put(0.0,0.)
  {
   \epsfig 
   {
   file=fig2_Lower_Left_hot_rasterPlot.eps,
   width=0.45\unitlength,
   }
  }
  \put(0.5,0.)
  {
   \epsfig
   {
   file=fig2_Lower_Right_hot_Histogram.eps,
   width=0.45\unitlength,
   }
  }
\end{picture}
\caption{Left column: The horizontal axis represents the time in MCS, while the vertical axis identifies the different units. White (black) pixel in the matrix entry $(i,\,j)$ means that the $i$ unit had a value -1 (1) at MCS $j$. Right column: Histogram of the overlaps (see footnote 1) between the MC configurations.}
\label{figure2}
\end{center}
\end{figure}

Figure~\ref{figure2}, left panels, show representative time courses of the neural states for the low-temperature ($\beta = 1.3$, top) and high-temperature ($\beta = 0.83$, bottom) cases. Right panels show, for the two cases, the distribution of the overlaps between all configurations in the time series. 

\begin{figure}[h!]
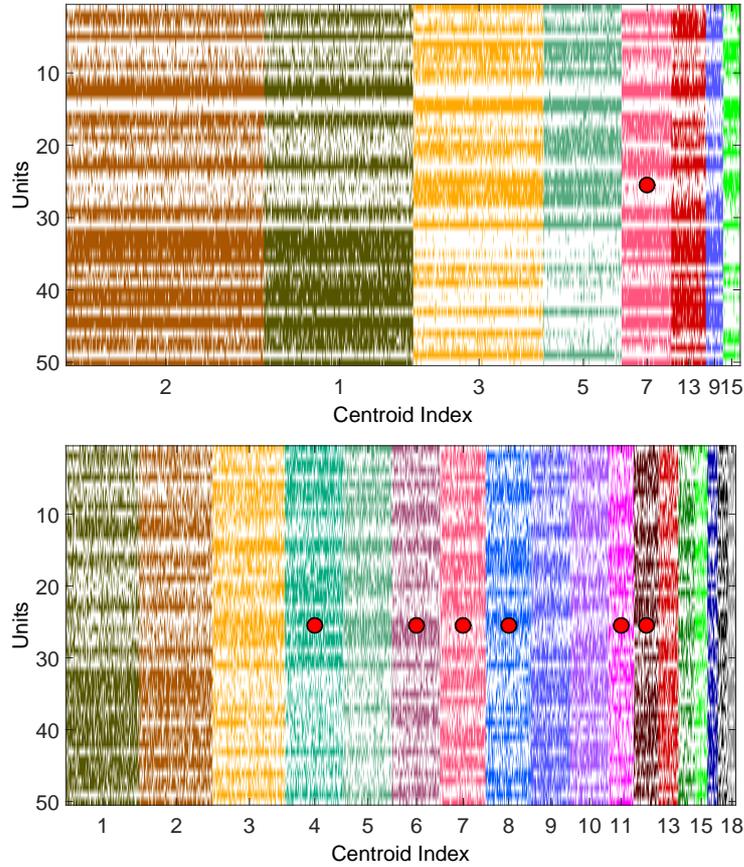

\begin{center}
\setlength{\unitlength}{\textwidth}
\begin{picture}(1,0.96)
\put(0.05,0.00)
  {
	\epsfig 
	{
	file=f3HotConfig.eps,
	width=0.8\unitlength,
	}
  }
  \put(0.05,0.48)
  {
	\epsfig 
	{
	file=f3ColdConfig.eps,
	width=0.8\unitlength,
	}
  }
\end{picture}
\caption{Clustered configurations for the raster plots in Fig.~\ref{figure2}. White is for units with value -1, while in colors (different for different clusters) are indicated units with value 1. The clusters whose centroids correspond to stored patterns or their reflections are marked by red dots. The centroids of the clusters appearing in the upper panel are a subset of those in the lower panel. Equal numbers refer to the same centroid. The clusters' masses can change due to finite size sampling, and clusters in each figure are ordered from biggest to smallest masses.}
\label{figure3}
\end{center}
\end{figure}

The subsequent Figure~\ref{figure3} illustrates the result of the clustering procedure for the low- and high-temperature cases. 

In the high-temperature case it is very difficult to discern a structure in the raster plot, and considering also the unimodal overlap distribution,  broadly symmetric around zero, it seems a challenging case for a procedure aimed at extracting the energy minima. 

This temperature dependence is reflected in the distribution of overlaps between the visited states. We note that, if energy minima would mostly coincide with the stored (uncorrelated) patterns, the overlap distribution would be tri-modal, with a peak in zero and two symmetric peaks at high (positive and negative) overlap, which is not the case even for low temperature (top-right panel in Figure~\ref{figure2}), with the distribution of visited states having average overlap about $0.5$ in magnitude. For higher temperature (bottom-right panel in Figure~\ref{figure2}) the overlap distribution is approximately Gaussian.

From Fig.~\ref{figure3} it is seen that for low temperature most centroids are indeed different from the stored patterns; for high temperature more centroids are identified, including the stored patterns.

While we did not perform a thorough analysis of clusters' centroids in terms of energy local minima, we checked that clusters are indeed akin to attractor basins, by measuring the configurations flow at zero temperature: starting the deterministic (zero-temperature Monte Carlo) dynamics from each one of the configurations assigned to a given cluster, the fraction of them for which the final overlap with their assigned centroid is larger than the initial one is $0.90 \pm 0.04$ (for the clusters found at $\beta = 0.83$), and $0.85 \pm 0.05$ (for the clusters found at $\beta = 1.3$).

\subsection{Clustering-aided Inference of Synaptic Connectivity}
\label{sec:results-inference}
\label{section4}
As mentioned in the Section Section \ref{sec:intro}, several methods have been and are being developed to infer effective synaptic connectivities from the time series of simultaneous neural recordings, some of which fall in the category of so-called ``Inverse Ising'' problems \cite{schneidman2006weak,roudi2015multi}.

Obvious hindrance in the application of such methods is the large number of parameters to be inferred (effective synaptic efficacies, of order $N^{2}$ for $N$ neurons), which makes them sensitive to noise in the data.

In the following, we show that the proposed clustering method can be used to formulate the inference problem in a reduced parameter space, using the activity configurations that are the identified centroids to parametrize a coupling matrix (inspired to the construction of the Hopfield connectivity matrix) with a number of parameters equal to the number of centroids.

Specifically (see also Section \ref{sec:methods}), the inference model has the form:
\begin{equation}
H[\boldsymbol{\sigma}]=-\frac{1}{N} \, \sum_{ij = 1}^N \sigma_i \, J_{ij} \, \sigma_j .
 \label{Ising}
\end{equation}

\noindent where the effective coupling matrix $J$ is the sum of weighted Hopfield-like terms $c_i^{\mu} c_j^{\mu}$ (see Eq.~\ref{eq4})\footnote{In the following, depending on the context, $\sigma_i$ will be either the activity of a binary neuron or a suitable binarization of the average activity of a population of spiking neurons.}:
\begin{equation}
J_{i j}=\frac{1}{N}\sum_{\mu=1}^C \omega_{\mu} c_i^{\mu} c_j^{\mu},
\label{eq6}
\end{equation}
\noindent $\boldsymbol{c}^\mu{}$ being the $C$ centroids identified by the clustering procedure, and the weights $\omega_{\mu}$ are to be inferred.

To test this approach, we first use again the Hopfield model data of Fig.~\ref{figure3}, bottom panel. Since 14 out of the 18 identified centroids are pairs of reflected patterns and give rise to the same Hopfield-like term in Eq.~\ref{eq6}, the number of parameters $\omega{}$ to be estimated is reduced to 11.

\begin{figure}[h!]
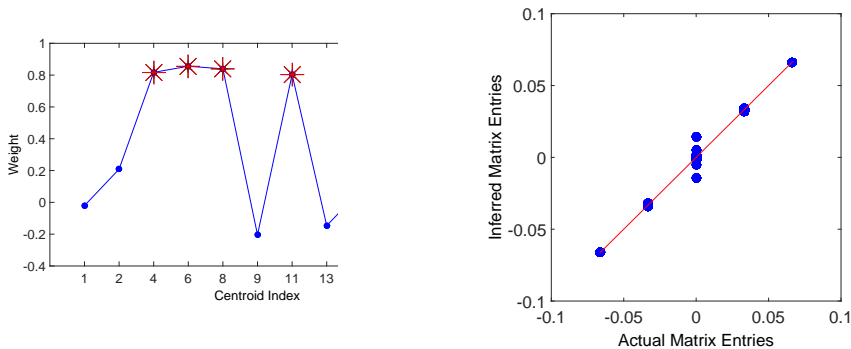

\begin{center}
\setlength{\unitlength}{\textwidth}
\begin{picture}(1,0.4)
  \put(-0.01,0.05)
  {
	\epsfig 
	{
	file=weights_centroidsDataSeriesA_asterisks.eps,
	width=0.5\unitlength,
	}
  }
   \put(0.51,0.00)
  {
	\epsfig 
	{
	file=DataSeriesA_Matrix_Inference_11C.eps,
	width=0.4\unitlength,
	}
  }
\end{picture}
\caption{Left: Inferred values of the weights $\omega_{\alpha}$ when fitting the reduced model (Eq.~\ref{eq6}) to the raster plot shown in the lower column of Fig.~\ref{figure2}). The red stars mark the weights corresponding to the patterns actually stored in the system. Right: Scatter plot of the inferred (reduced model) synaptic couplings against the actual ones. The red curve shows the identity as a reference}
\label{figure4}
\end{center}
\end{figure}

In Figure~\ref{figure4}, left panel,  we show the inferred values of $\omega_{\mu}$. Only four of them (marked with an asterisk) are significantly different from zero, and they correspond to the centroids that coincide with the $4$ stored patterns. 

Notice that, when the $\omega{}$s are roughly equal, they play a role similar to an inverse temperature; indeed, the value of $\beta$ used to generate the time series (0.83) is very close to the $\omega$ values for the centroids corresponding to the patterns.

We also remark that the obtained values for $\omega$ would not be trivially expected from the structure of the state space, in that not only 7 out of 11 centroids (not counting reflections) do not belong to the stored patterns, but the clusters of largest mass are not centered on patterns (remember that the Hopfield system is far from the retrieval phase).

\begin{figure}[h!]
\begin{center}
\setlength{\unitlength}{\textwidth}
\begin{picture}(0.61,0.46)
  \put(0.0,0.00)
  {
	\epsfig 
	{
	file=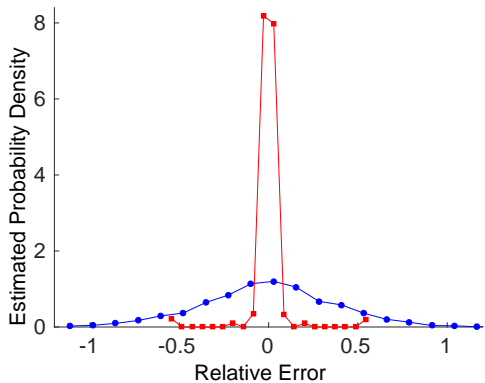,
	width=0.6\unitlength,
	}
  }
\end{picture}
\caption{A posteriori probabilities of inference errors for the cases of the reduced model (red curve) and the full matrix inference (blue curve). Details in the text.}
\label{figure6}
\end{center}
\end{figure}

In Figure~\ref{figure4}, right panel, we compare the inferred $J$  with the real $J$ ($N \times (N - 1) = 2450$ elements, taking only $P + 1 = 5$ values); the inference is actually good (the continuous identity line is drawn to guide the eye), and better than the one obtained inferring directly the full $J$ matrix (i.e. not adopting the parametrization of the synaptic matrix in Eq.~\ref{eq6}), as shown in Fig.~\ref{figure6}, where we compare the relative error ($(J_{ij}^{Inferred} - J_{ij}) / \langle |J_{ij}| \rangle_{ij}$) for the two cases.

From this simple example, we confirm that the adopted parametrization, besides the obvious greater simplicity and lesser computational load, is effective in reducing the effect of noise in the data and makes the inference less prone to overfitting, while allowing a good match between model and data.

\subsection{Towards a more realistic scenario: a multi-modular network of spiking neurons matching a prescribed spatial correlation structure}
\label{sec:results-spiking}
\label{section5}

In the previous Section we showed that, from the spatial correlation structure of the configurations sampled by Monte Carlo, MS clustering was effective in reconstructing the main local energy minima, and allowed for a parsimonious parametrization of the synaptic matrix that afforded better inference. In that case, the correspondence between local density maxima in the state space, and local energy minima, was ensured by the existence of a Gibbs equilibrium probability for the Hopfield model.

In the perspective of applicability to real electrophysiology data, in the present Section we extend the approach to networks of spiking neurons, and to a situation where the notion of a static energy landscape is no longer strictly applicable.

To establish a meaningful benchmark, we want to preserve some a priori knowledge of key features of the state space, to be checked against the found cluster structure, and we do this by setting up a spiking network constructed so as to possess a prescribed spatial correlation structure, from which we generate time series to be clustered. To achieve this we develop a method that, we believe, has a wider interest beyond the case at hand.

The chosen network architecture is composed of strongly self-coupled modules of spiking (integrate-and-fire) neurons, with much ($10^{-2}-10^{-3}$) weaker synaptic couplings between modules. Intra-module synapses are drawn from a Gaussian distribution, with mean and average chosen (using mean-field predictions) such that each module in isolation is approximately bistable, between states of low (DOWN) and high (UP) firing activity. 

The interest in this choice for the architecture of the spiking network stems from accumulating evidence that, not only a modular structure is suggested by the mesoscopic anatomical organization of the cortex, but it also appears to be recognizable in the cortical neural dynamics, which can proceed as the dynamic composition of abrupt jumps between UP and DOWN states (\cite{MattiaEtAlJNeurosci2013,latimer2015single}), as mentioned in the Introduction (see also \cite{abeles1995cortical}\cite{Luczak2007}\cite{mazzucato2015dynamics}).

We remark that synapses are not be constrained to be symmetric, therefore the clustering procedure will not, strictly speaking, match the minima of an energy function of the system. Besides, the Integrate-and-fire spiking neuron model (see Section \ref{sec:methods}) is endowed with spike-frequency adaptation (SFA), a much studied (and pervasively observed) self-inhibitory mechanism depending on the activity history of the neuron; SFA makes the effective landscape, even when it exists, locally dynamic at the point currently corresponding to the network state. 

Fig.~\ref{figure7} describes the main steps involved in the network construction. The multi-modular network is sketched (panel A) as a collection of $64$ neural modules, each composed of $32$ adapting excitatory and $16$ inhibitory neurons (see see Section \ref{sec:methods} for details). The approximate bistability of the single modules, which is to a large extent preserved in the interacting network, is illustrated in panel B by the time course of the firing rate of the excitatory neurons from two sample modules; the resulting bimodal distribution of firing rates allows to binarize the modules' activity, as shown in panel C \footnote{We remark that the binarization step, here instrumental for the construction of the intended benchmark, would not be needed in general, and spate-space clustering can be performed of the raw multi-dimensional time series.}. Once binarized, the time course of the whole network activity can be represented as a sequence of binary vectors (see the `raster plot' in panel D), to which clustering is applied.

\begin{figure}[h!]
\setlength{\unitlength}{\textwidth}
\begin{picture}(1.1,0.8)
\put(-0.1,0.0)
  {
	\epsfig 
	{
	file=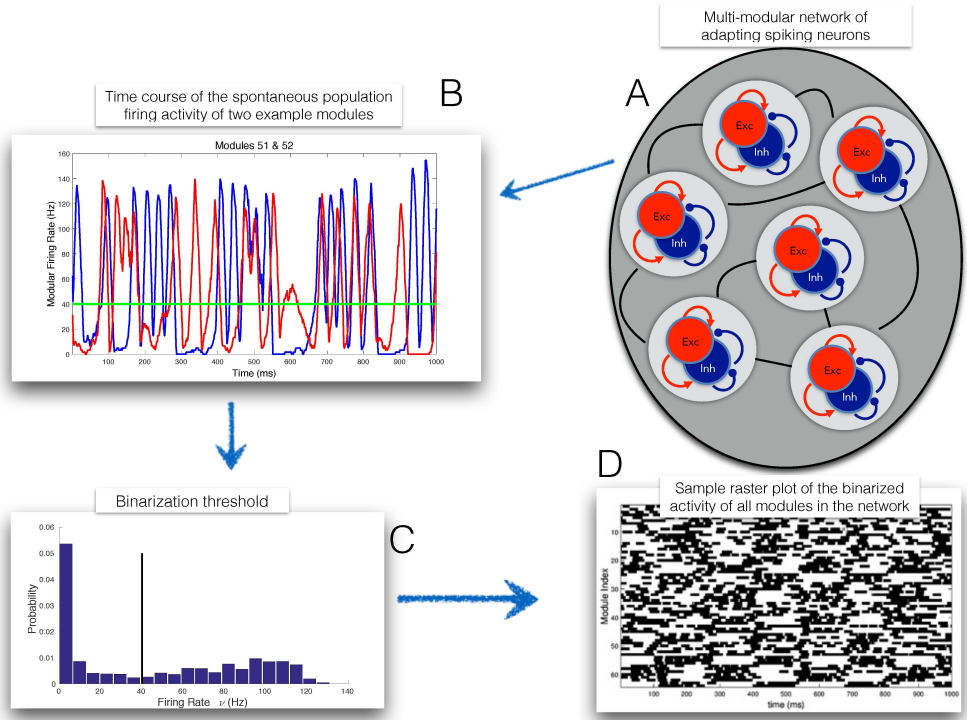,
	width=1.1\unitlength,
	}
  }
\end{picture}
\caption{Sketch of the main steps involved in the network construction. See text for details.}
\label{figure7}
\end{figure}

We choose to assign a spatial correlation structure mimicking the one of a Hopfield attractor network, each module corresponding to one binary neuron of the Hopfield network: inter-module synapses, initially drawn from a Gaussian distribution, are subject to an iterative procedure (see below, and Section \ref{sec:methods} for details) to match those of the reference Hopfield model. 

Were it not for the effects of SFA, the correlation-matched multi-modular network would be expected to behave as an attractor, Hopfield-like system, with the Up and Down states of each module playing the role of the binary values of the Hopfield neurons. However, as the network enters the basin of one attractor, the active modules start lowering their activity because of SFA, thereby destabilizing the state: as noted above, the attractor landscape becomes `dynamic', in that its depth and curvature around attractor states get lower when the system visits them, promoting transitions to other basins, which are biased by the correlations with other attractor states (due to finite-size effects) \footnote{The role of SFA in promoting transitions between neural states has been studied in various contexts, see {\it e.g.} \cite{abdollah2012optimal,duarte2014dynamic,huguet2014noise,roach2016memory,deco2005sequential,theodoni2011neuronal,mattia2012exploring,gigante2007diverse,akrami2012lateral}.}.

The new procedure we developed to determine the inter-module synaptic couplings is inspired to the Boltzmann learning strategy where, at each iteration, the change in the couplings is proportional to the difference between the spatial correlations in the model and in the data (and analogously for the external fields). For Boltzmann learning, such difference is the gradient of the Kullback-Leibler distance between the state probabilities in the model and in the data. In our spiking network the explicit form of such a function is lacking; still, intuition suggests, and we assume, that a monotonic relation still holds between the synaptic couplings and the spatial correlations, and this is equivalent to assume that we know the sign of an unknown gradient. Formulated in this way, our `pseudo-Boltzmann' iterative process to find the optimal synaptic couplings is naturally mapped onto the Rprop algorithm (see Section \ref{sec:methods}, Eqs.~\ref{eqRprop1}, \ref{eqRprop2}).

In summary, the sequence of steps we take is the following: 1) store a set of patterns {\it a la} Hopfield in a network of binary neurons (as in the previous section, the Hopfield network will be in its glassy phase); 2) from the Hopfield network, measure spatial correlations and site magnetizations (average activity); 3) set up a multi-modular spiking network of approximately bistable modules; 4) use pseudo-Boltzmann learning to find inter-modular couplings and external rates to mimic correlations and magnetizations of the Hopfield system; 5) perform MS clustering on the configurations generated by simulations of the spiking multi-modular system; 6) check the quality of the result (see below) .

The chosen `reference' Hopfield model has $N=64$, $P=4$, $\beta=1.1$, for which we generate a long Monte Carlo sequence, and measure the spatial correlations and site magnetization that are to be matched by the multi-modular spiking network with optimal inter-module couplings $J_{pq}$ and external inputs $\nu_p^{ext}$ obtained from the pseudo-Boltzmann procedure explained above.

One may ask whether the simplest choice that intuition would suggest, {\it i.e.} simply taking the inter-module synapses as proportional to the computed Hopfield ones, would work. We checked this option, with poor results. One reason is that the low- and high-firing rate states (which are subject to binarization in the clustering procedure) are not dynamically equivalent (contrary to the binary neurons of the Hopfield case): one state can be deterministically more stable than the other ({\it e.g.} in the sense of the linear stability of the two corresponding fixed points in the mean-field approximation); furthermore, noise is higher in the high-firing rate state (finite-size noise is activity-dependent and acts multiplicatively). Finally, there are `quenched noise' effects: the synaptic connectivity in each module is a different (small) realization of the same probabilistic model, and can lead to quite different dynamics between modules.

We checked the success of pseudo-Boltzmann learning in several ways.

First, the correlation between the partial correlations measured from the optimal spiking network and those from the reference Hopfield model is very high ($R^{2} = 0.975$); the average `magnetization' for the optimal spiking network is order $10^{-3}$, to be compared with the potential range $[-1,1]$ and the theoretical null value for the Hopfield network. The found optimal external firing rates have large variations between modules (over $\pm 20 \%$ with respect to their average), which is expected, since they contribute to compensate for the heterogeneous excitability of the different modules.

Second, we checked the similarity between the cluster structures emerging from the optimal spiking network and that of the reference Hopfield network, by computing the absolute value $|q|$ of the overlaps between the centroids found for the two networks ($c^{S}$ and $c^{H}$ respectively for the spiking and Hopfield networks).

Out of the 12 $c^{H}$ centroids, $9$ of them ($75\%$) have $|q|=1$ with at least one of the $c^{S}$; among the others, 2 of them have $|q| \sim 0.97$ with at least one of the $c^{S}$; 1 of them has $|q| \sim 0.84$ with one of the $c^{S}$.

Conversely, out of the 13 $c^{S}$ centroids, 8 of them (62\%) have $|q|=1$ with at least one of the $c^{H}$; among the others,
2 of them have $|q| \sim 0.97$ with at least one of the $c^{S}$; 3 of them have $|q| \sim 0.75$ with at least one of the $c^{H}$.

This shows that the pseudo-Boltzmann iterative procedure, by enforcing approximately equal mean activities and spatial pair correlations between the Hopfield model and the modules of the spiking network results in fact in similar mostly visit regions of the state space for the two systems.

Third, we inferred inter-modular synaptic efficacies (using MPF as in the previous Section) from the spiking network time series, and found that for all modules pairs the inferred synapses are close to corresponding synapses of the reference Hopfield model (the mean absolute value of the relative error, taken over the $J$ entries, is $11 \%$, likely mostly due to finite-sample noise, as suggested by comparison with Fig.~\ref{figure6}); this confirms that, despite the large differences between the synapses of the Hopfield network and the optimal  inter-modular average synaptic efficacies determined by pseudo-Boltzmann learning (not shown), the dynamics of the resulting spiking network effectively embodies inter-module interactions consistent with the reference Hopfield network.

To summarize, the somewhat complex procedure described allowed us to construct a modular spiking system with some control on desired features of the state space, not easily enforced by simple {\it ad hoc} assignment of the synaptic structure; while in this case the construction was guided by a reference Hopfield model, whose neurons were naturally mapped onto the network's modules, more in general we believe the procedure is interesting {\it per se}, as a means to enforce a prescribed pattern of spatial correlations (and associated state-space structure) in relatively complex networks of spiking neurons.

\subsection{Dynamics in the centroid space}
\label{sec:results-lz}
\label{section6}

When performing MS clustering in the state space, information about the dynamics of the original time series is - by construction - lost. However, once clustering is done, knowledge of the clusters' centroids allows to go back to the multi-dimensional dynamics, and cast it in a useful compact form: to each one of the vectors expressing the states at successive sampling times, we substitute the label of the cluster that vector was assigned to. 

The description of the system's dynamics is reduced to a `symbolic dynamics' in the centroids space, which opens up options to expose dynamic features that may be difficult to uncover directly from the analysis of the spiking activity, as we illustrate through an example in the present section.

Based on the procedure developed in the previous Section to set up multi-modular spiking networks, we want to generate a family of networks for which different degrees of `complexity' can be expected, instantiated here in different dynamic memory span, induced by the SFA component that, as discussed, affects locally the dynamics in a history-dependent way.

In the previous section SFA was chosen small, just enough to obtain measurable state transition rates in the simulation time. 
Here we set up a series of spiking multi-modular networks, each one constructed as in the previous section, but with different parameters for SFA. 

We span a range of values for the timescale of SFA ($\tau_{SFA}$), while keeping the product $\tau_{SFA} \, g_{SFA}$ constant (in order to keep the SFA `strength' constant and have comparable systems, see Section \ref{sec:methods}). 

We show here that the reduced dynamics in the centroid space lends itself naturally to methods for symbols-oriented measures of complexity, able to easily capture memory effects. 

It has long been suggested, and reported in several published works, that the Lempel-Ziv (LZ) complexity measure \cite{lempel1976complexity} may be usefully adapted to characterize neural data series \cite{kaspar1987easily,abasolo2015lempel,amigo2004estimating}; in particular, a suitably normalized LZ complexity has been successfully employed as a diagnostic measure of the distance to the conscious state in neurological patients \cite{casali2013theoretically}. 
In essence, the approach is based on the intuitive idea that the more complex the signal, the less its compressibility; in other words, more structure in the signal increases its predictability. Therefore, a memoryless stochastic time series would have maximal complexity, and any memory embedded in the dynamics generating the time series would make it decrease. Such entropic measures provide information beyond what linear correlation analysis can provide.

We therefore measure, for the symbolic sequence reduction of the multi-dimensional dynamics obtained from networks with different $\tau_{SFA}$, LZ complexity and a relative complexity index, as detailed in Section \ref{sec:methods}.

Expectation is that increasing $\tau_{SFA}$ generates multidimensional time series with decreasing complexity. We checked that such expectation is met in our spiking simulation data, and that differences in LZ complexity for high and low $\tau_{SFA}$ are statistically significant for the simulated time span. 

In order to quantify the difference, for each value of $\tau_{SFA}$ and $g_{SFA}$ we also simulated ten realizations of Markov chains in the centroid space, generated by the transition probabilities estimated from the simulation, thereby producing surrogate memoryless sequences with the same transition statistics as the data; we computed the LZ complexity averaged over the ten Markov chains (`$c_{\mathrm{Markov}}$'), and to compare it to the LZ complexity from the simulation data (`$c_{\mathrm{sample}}$'), we computed the ratio $R=(c_{\mathrm{Markov}} - c_{\mathrm{sample}})/ c_{\mathrm{Markov}}$ (see Section \ref{sec:methods}), \footnote{including self-transitions would blur the difference and decrease the LZ complexity, since the corresponding runs of identical centroid labels add to the compressibility of the sequence; we did not consider self-transitions in the analysis}. 

For a meaningful comparative analysis of the complexity of the dynamics in the centroid space for widely different values of $\tau_{SFA}$ (and related $g_{SFA}$), we should make sure that the corresponding networks are indeed similar between themselves, and with the reference Hopfield network, in terms of the state space structure.

For this purpose we did a preliminary analysis of the clusters found for all the explored values of $\tau_{SFA}$; since, as $\tau_{SFA}$ increases, an increasing number of small clusters appears\footnote{This is due in part to the fact that, because of SFA, `bridge states' appear, with associated small clusters, where the system transits just after SFA has destabilized one major attractor state, on its way to make a transition to another major attractor state.}, with small differences from main ones, we first clustered the centroids with standard methods (hierarchical clustering using Hamming distance and complete linkage), to obtain a `fuzzy' version of the whole set of main centroids; for those (76 clusters), we found that:
the first $8$ fuzzy centroids account for $66.3\%$ of the total configuration mass and are detected, on average, for $84.0\%$ of the $\tau_{SFA}$ values; the first $10$ fuzzy centroids account for $71.7\%$ of the total configuration mass and are detected, on average, for $73.0\%$ of the $\tau_{SFA}$ values; the centroids of the reference Hopfield network are recovered in $87.2\%$ of cases (by `recovered' we mean that at least one of the centroids found for an Hopfield simulation-clustering has overlap $> 0.719$ with the considered Hopfield centroid, the threshold value $0.719$ being determined such that $95\%$ of the overlaps between the centroids found for the considered case are below such threshold.

The results of the analysis are summarized in Figure~\ref{figure7}: we see that the ratio $R$ increases with increasing $\tau_{SFA}$ or, in other words, that as memory effects increase the complexity of the actual centroid sequences gets increasingly larger than that of surrogate Markov sequences, which provides a quantitative information on the non-Markovian nature\footnote{We use the term here in an informal sense, not distinguishing between higher-order Markovian and strictly non-Markovian processes.}) of dynamics for higher $\tau_{SFA}$ cases. In the figure we also report for comparison (red line) the $R$ value obtained for the centroid sequence extracted from the time series generated by the reference Hopfield network (which is inherently Markovian); it is seen that, although the centroid sequence deviates from a Markov process, the relative difference w.r.t. the surrogate sequence is very small, much smaller than the one for the spiking dynamics, where SFA plays a major role.

\begin{figure}[h!]
\setlength{\unitlength}{\textwidth}
\begin{picture}(1,0.53)
\put(0.05,0.0)
  {
	\epsfig 
	{
	file=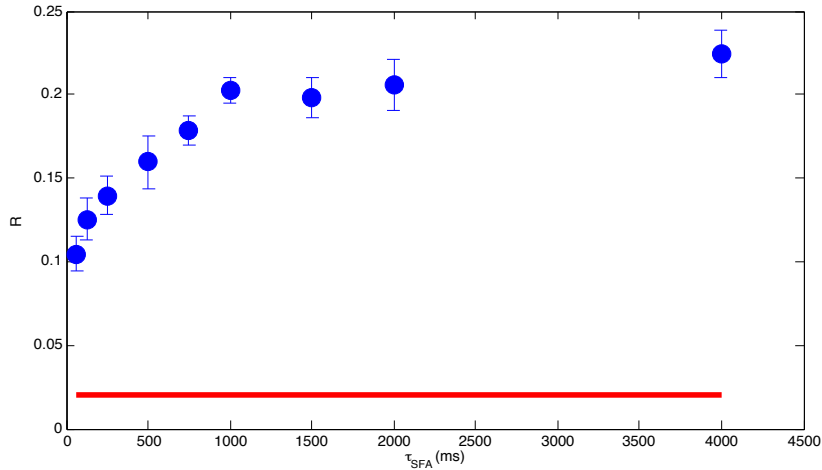,
	width=0.9\unitlength
	}
 }
\end{picture}	
\caption{Dependence of the complexity measure $R$ {\it vs} $\tau_{SFA}$. Red line: $R$ computed for the Markovian sequence generated by the reference Hopfield system; blue line: $R$ {\it vs} $\tau_{SFA}$ for the actual sequence.}
\label{figure7}
\end{figure}

Of course, deviations from a Markov, memoryless dynamics can take many forms; a generic expectation is that, for a non-Markov process, sequences of states of given length are more, or less, likely to occur than predicted based only on the transition probability matrix. 

To gain insight for the case at hand, we considered the centroid sequence for the spiking network with the highest $\tau_{SFA} = 4$s (of length about $1.3 \times10^{3}$). For each of the triplets of centroid labels ($13^3$ triplets, since $13$ centroids were extracted in this case by MS clustering) the probability of occurrence was estimated from the actual sample, and computed from the Markov transition probabilities estimated from the same sample. 

\begin{figure}[h!]
\setlength{\unitlength}{\textwidth}
\begin{picture}(1.0,0.50)
\put(0.0,0.0)
  {
	\epsfig
   {
   file=TripletProbabilitiesTau4000.eps,
   width=0.5\unitlength
   }
 }
 \put(0.51,-0.125)
  {
      \epsfig
	{
	file=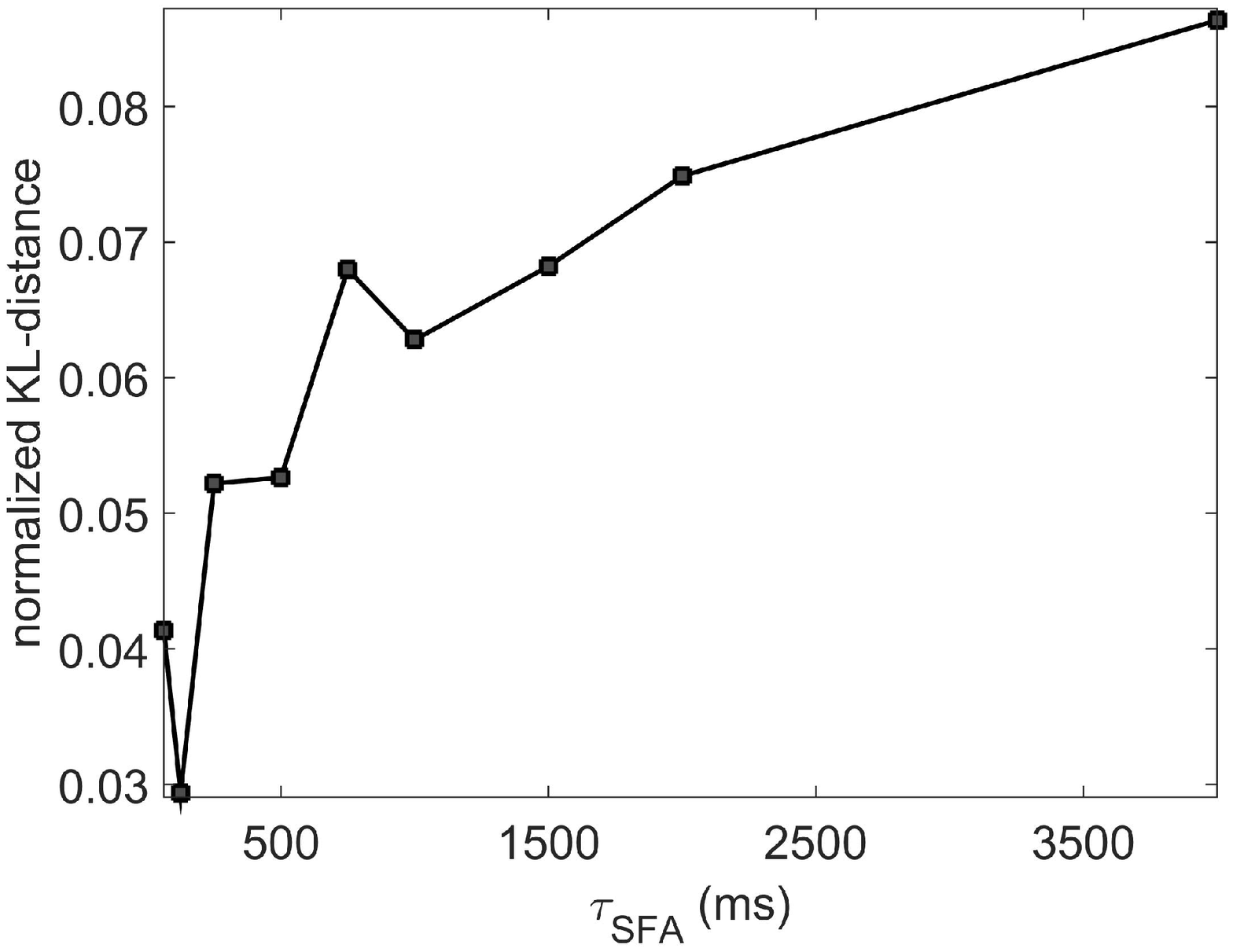,
	width=0.49\unitlength
	}
  }
\end{picture}	
\caption{Left: for all triplets occurring more than $10$ times, the blue points are the estimated probability from the actual sequence {\it vs} the computed Markov probabilities from the estimated transition probability matrix; green crosses are the estimated probability from the actual sequence {\it vs} the estimated probability from the Markov surrogates. Right: the Kullback-Leibler distance between the sampled triplets distributions from the actual sequence and the surrogate Markov sequence, as a function of $\tau_{SFA}$.}
\label{figure8}
\end{figure}

Fig.~\ref{figure8}, left panel, shows (blue points) a scatter plot of such probabilities (limited to the triplets occurring more than $10$ times). With reference to the black identity line (close to which the points would obviously cluster if the actual sequence was Markovian), it is seen that many triplets are over- or under-represented in the actual sequence (particularly in the region of higher probabilities which matters most), as a reflection of its non-Markovian nature.

In order to assess the significance of the observed differences, we also generated a truly Markovian sequence with the same transition probabilities and of the same length as the sample, and from it we estimated the triplets probabilities; the green crosses show the corresponding scatter plot with the computed Markov probabilities, and it is clearly seen that finite-sample effects are much smaller than the spread observed for the actual sequence, confirming its genuine non-Markovian nature.

Finally, given the dependence of $R$ on $\tau_{SFA}$ shown in Fig.~\ref{figure7}, it was natural to ask how the non-Markovian estimated triplets occurrence probabilities would depend on $\tau_{SFA}$; this is illustrated in Fig.~\ref{figure8}, right panel, where we report the Kullback-Leibler distance between the sampled triplets distributions from the actual sequence and the surrogate Markov sequence, as a function of $\tau_{SFA}$; although in principle for different $\tau_{SFA}$ we may expect different contributions to the non-Markov nature of the sequence from sub-sequences of different length, we observe an approximately monotonic dependence of the KL distance on $\tau_{SFA}$.

In summary, reducing the multidimensional time series of spiking data to the sequence of labels of the clusters identified by the state-space clustering procedure, casts the dynamics in a form easily suited to capture and quantify traces of memory effects, for instance allowing in principle the comparison between recordings in different experimental conditions.

\section{Discussion}
\label{sec:discussion}
\label{discussion}
 
In this work we considered pre-existing strategies and developed new tools that, taken together, compose a methodology with good potential, we believe, in analysis and modeling of neuroscience data.

We started from a simple idea (though, to our knowledge, it was not exploited so far in the analysis of multiple neural recordings): to represent the multidimensional time series of neural activities as a density distribution in a corresponding multidimensional space, and perform a density-based clustering procedure to extract the local density maxima. We strived to show that this type of dimensional reduction is in fact a versatile instrument; in particular we illustrated examples that it can be useful in achieving better inference of synaptic couplings from neural activities, and also to cast the multi-dimensional neural dynamics in a compact form amenable to symbol-oriented methods of complexity analysis.

The clusters, and the associated centroids, have an obvious interpretation when the original time series is generated by an attractor dynamics, but retain an informative value even when (like in the case of SFA in the spiking network) the picture of a static attractor landscape is no longer appropriate; preliminary work in progress on multiple in-vivo recordings during motor tasks makes us confident that the approach can provide compact and usable information even in strongly non-stationary conditions.

We first validated the method on the activity generated by a simple Hopfield network, albeit we choose for it a highly nontrivial working regime in which the energy landscape explored during the network dynamics is `glassy', with many local minima uncorrelated from the memory pattern embedded in the Hebbian synaptic matrix. We showed that our (modified) Mean-Shift clustering is indeed effective in revealing features of the energy landscape from very noisy time series. Besides, knowledge of the clusters' centroids allowed an efficient parametrization of the synaptic matrix, which in turn allowed a much better result when inferring the synaptic efficacies from the network's sampled activity.

We should remark that the generality and robustness of this latter result is still to be thoroughly investigated; in the case shown here, the mathematical form of the inference model is similar (though not equal) to the one of the model network generating the data used for inference. While this is indeed a limitation shared with most published works dealing with `inverse Ising' approaches to synaptic inference, in future work we want to systematically explore more generic forms of parametrization of the synaptic matrix (still using information from the clusters extracted from the activity time series).

Furthermore, the proposed approach addresses, though admittedly in a specific context, a general issue.
Whatever the specific method adopted, the extent to which the inference of single synapses can be trusted can be severely affected by several factors, like inherent inadequacy of the model used for inference, poor quality of the data and noise, limited data sample. Still, it is usually (explicitly or implicitly) assumed that even when the inference procedure fails to match single synaptic efficacies, if the synaptic matrix has a global structure it should still be captured in the inferred matrix. This would almost inevitably call for some kind of dimensional reduction of the inferred synaptic matrix, such that the informative relevant `mesoscopic' structure is retained. In a sense, what we propose here can be viewed as a way to formulate a `mesoscopic' inference problem in the first place; again, we cannot claim full generality here, but our clear success in the case of attractor networks examined in the present work makes the approach, we believe, an interesting option to be further explored.

Wanting to move to more realistic network models, we were naturally led to networks of spiking neurons (Integrate-and-fire, with SFA); in the Introduction and in the Results we provided motivations for a specific choice of the network architecture as composed of weakly coupled, individually bistable neural populations. In order to make contact with the study of clustering in the Hopfield network and, more importantly, to have some {\it a priori} knowledge of the effective landscape of the spiking system, we also wanted to set up the inter-module couplings such that the spiking network as a whole would share static properties of the state space with an Hopfield attractor network. This need motivated us to develop a procedure that determines the inter-modular couplings such that the network's activity generates a prescribed pattern of pairwise spatial correlations.

The procedure involves a new algorithm that extends the domain of applicability of Boltzmann learning, and uses Rprop learning; as already remarked, we believe this approach has a value beyond the specific purpose it served in the present work. 

It rests on the intuition that between excitatory synaptic connection strength and neuronal activity correlation a monotone relationship should hold. 
On a more general level, this is just an instance of a strategy (which is also a human ability) to identify the relevant variables in a problem and code them in such a way that they have a conditionally monotone relationship with the relevant observables or, more specifically, with the statistic deemed sufficient for the problem under exam \cite{dawes1979robust}. Such ability, coupled with the proven effectiveness in a variety of contexts of optimization algorithms that depend only on the sign of the derivative of the function to be optimized (as Rprop)\cite{Rprop,igel2003empirical}, can make the idea behind the proposed algorithm robustly generalizable to a wide array of problems, dealing both with static and dynamic properties of neuronal networks, \textit{e.g.} by taking into account spatial as well as temporal correlations.

Finally we went back to dynamics. A natural step was to substitute the original multidimensional time series with the corresponding `symbolic dynamics' of centroid labels, and to ask whether the resulting paths in the centroid space would allow to extract information on the system's original dynamics that would be difficult to directly expose. A case in point, in our context, was to inspect how the sequences of transitions between centroids in the spiking modular network would depend on the strength of SFA. SFA introduces `memory' in the dynamics, and higher SFA makes the original time series more history-dependent, and the corresponding symbolic dynamics of centroid labels is expected to be less Markovian. As discussed in the text, the pattern of transitions between centroids results from an interplay, in the original time series, between noise, spatial overlaps between attractor states of the multi-modular network, and SFA-dependent effects.

To quantify the memory-related complexity of the network activity, we defined a measure based on Lempel-Ziv (LZ) complexity (inspired to previous work in various scientific domains, including neuroscience): for different time scales of SFA, we compared the LZ complexity of the centroid time series with surrogate Markov sequences with the same transition probabilities, finding that, as expected, longer timescales of SFA correspond to less complex (and less Markovian) centroid sequences. We also provided insight into such SFA-dependent non-Markovian nature by studying the occurrence probability of selected sub-sequences.

Again, while we illustrated in some detail this reduction to symbolic dynamics and the analysis of its complexity in the specific case under consideration, its value rests with its generic applicability to multidimensional time series.
 
A few closing remarks, to facilitate comparison with other approaches to the characterization of multidimensional time series in neuroscience (including Hidden Markov Models, that recently have been frequently used in the analysis of neural data (see {e.g.})). 

First, the proposed state-space approach is free from bias towards spherical clusters and from a pre-defined number of clusters. Such freedom is inherent in the approach taken here, which is also quite easy to implement.

Second, for large data sets the density-based clustering approach can become computationally expensive, and of course it is meaningful to try to speed it up. A recent successful attempt was made in \cite{rodriguez2014clustering}, where a preliminary distance-based selection procedure excludes outliers (points with low local density) from the iterative procedure. The validity and performance of the approach is tested in a variety of benchmark data sets, including the Olivetti face dataset \cite{Olivetti}.
In comparing to the present work, we remark that on the one hand we did not make an effort towards computational efficiency; rather, we wanted to show the potential of density-based clustering for the analysis of neural data from multiple simultaneous recordings. On the other hand, we tried the method proposed in \cite{rodriguez2014clustering}, and checked that it performs poorly in several representative situations, due to the high level of noise. A reasonable strategy would probably be a mixed method which first performs a number of iterations (dependent on the noise level) of the mean-shift algorithm (to `clean up' enough the data distribution in the configuration space), followed by the faster procedure described in \cite{rodriguez2014clustering}.

\section{Materials and Methods}
\label{sec:methods}

\subsection{Clustering in the state space: a modified Mean Shift Algorithm}
\label{sec:meanshift}

\label{section2}
We start with a data set $\{x_1, \, x_2, \, \dots x_M\}$ consisting of $M$ (real valued) vectors in a $N$-dimensional space. The Mean Shift (MS) algorithm [Fukunaga Hostetler]\cite{Fukunaga} iteratively picks a random $i$ ($1 \leq i \leq M$) and at step $t$ updates the vector $x_i$ according to the rule:
\begin{equation}
\label{eq.MeanShift}
x_i^{t + 1} = \frac{\sum_{j \neq i}^{1,M} x_j^t \, w\Big( \mathrm{dist}(x_i^t, \, x_j^t) \Big)}{\sum_{j\neq i}^{1,M} w\Big( \mathrm{dist}(x_i^t, \, x_j^t) \Big)};
\end{equation}
where $w(\cdot)$ is a non increasing function of its argument, $\mathrm{dist}(x_i^t, \, x_j^t)$ is a distance measure between vectors, and the original vector set is $\{ x_i^0 \}$. In other words, at each step $x_i^t$ is replaced by a weighted average of the other points $x_j^t$: points that are closer to $x_i^t$ are given more weight, thus defining a ``soft'' neighborhood of $x_i^t$ (see Fig.~\ref{figure2}). Therefore the update rule effectively moves $x_i$ towards regions of the $N$ dimensional space where the (local) density of points is higher at step $t$. The other $M - 1$ vectors are instead left unchanged ($x_j^{t + 1} = x_j^t$ for $j \neq i$); though a parallel, deterministic update rule is in principle possible applying Eq.~\ref{eq.MeanShift} to all the data points at the same time, the random, sequential update just described proved more robust in practice.

The iteration terminates upon a threshold condition: when all $x_i(t)$ belong  to a discrete space, the iteration terminates when the fraction of cases (out of the last $M$) in which $x_i(t + 1) \neq x_i(t)$ is below a fixed threshold; in the continuous case, an additional threshold on a minimal displacement is required. Clusters are finally determined as sets of initial points $x_i^0$ absorbed by a same final point (their number we call the `mass' of the centroid); each final point can be taken as representative of a different cluster (its `centroid'.)

The choice of the weighting function $w(\cdot)$ is central to the MS algorithm. Such function can in general depend both on the step $t$ and the configuration of the neighborhood, thus adapting to different conditions. We used a step weight function:
\begin{equation}
w(x) = \left\{\begin{array}{@{}l@{}}
    1 \quad \mathrm{if} \; x \leq r \\
    0 \quad \mathrm{if} \; x > r
  \end{array}\right. .
\end{equation}

The choice of the radius $r$ is critical; on the one hand, a small radius can make the algorithm too sensitive to local variation of density (e.g. due to noisy data), leading to many non-representative clusters; on the other, as the radius increases, the reduced sensitivity to local fluctuations can lead to merging  different meaningful clusters. We therefore resorted to an adaptive heuristic criterion to select $r$. At each step $t$, we compute the standard deviation $\sigma(n)$ of the distances between $x_i^t$, the point selected for update, and its $n$ nearest neighbors. We then take $n_{\min} = \arg \min_{n} \sigma(n)$, and set $r$ as the distance between $x_i^{t}$ and its $n_{\min}$-th nearest neighbors. We found such heuristics to work well in our case of discrete (hyper cubic) space, for different data sets and for a wide distribution of cluster sizes.

This criterion for determining the radius, together with the additional MS clustering  performed on the centroids found in a first run (see Section \ref{sec:results}), constitute a modification of the original MS algorithm, which we found advantageous.

In the MS algorithm the initial ordering of data points is irrelevant, as they are treated as samples of a static distribution. For time series this means that the time structure of the data is completely ignored and points that are distant in time can end up being clustered together, \textit{e.g.} as the result of subsequent `jumps' of the system in a same region of the state-space. Nonetheless, as we saw in Section \ref{sec:results-lz}, segmenting the original time series in terms of transitions from one cluster to another can highlight relevant dynamical features (non-stationarities, memory effects) in the data.

\subsection{Inference}
\label{sec:inference}
Specifically, we will assume for the system under consideration an effective Ising-like energy function:
\begin{equation}
H[\boldsymbol{\sigma}]=-\frac{1}{N} \, \sum_{ij = 1}^N \sigma_i \, J_{ij} \, \sigma_j .
 \label{Ising}
\end{equation}

We propose to decompose the effective coupling matrix $J$ in a set of weighted Hopfield-like terms $c_i^{\mu} c_j^{\mu}$ (see Eq.~\ref{eq4}):
\begin{equation}
J_{i j}=\frac{1}{N}\sum_{\mu=1}^C \omega_{\mu} c_i^{\mu} c_j^{\mu},
\label{eq6}
\end{equation}
\noindent where the vectors $\boldsymbol{c}^\mu{}$ are the $C$ centroids identified by the clustering procedure, and the weights $\omega_{\mu}$ are to be found through an optimization procedure.

Notice that, when the $\omega{}$s are roughly equal, they play a role similar to an inverse temperature; indeed, we found that the value of $\beta$ used in the Hopfield simulations is very close to the $\omega$ values for the centroids corresponding to the patterns.

For the optimization we use a recently introduced technique, Minimum Probability Flow (MPF) \cite{MPF}, with the advantage that it does not rely neither on Monte Carlo (MC) runs (like Boltzmann learning does) nor on mean field approximations (like most Inverse Ising methods do \cite{ClassicInverse}). It is based on the maximization of a function whose maxima are close to the Boltzmann Likelihood ones, but with no needs of performing computationally expensive Gibbs samplings.

Whilst Boltzmann learning aims to minimize the distance (more precisely the Kullback-Liebler divergence) between the distribution of the data and the distribution of the model, MPF aims to minimize the distance between the distribution of the data and the model distribution after an infinitesimal Monte Carlo step, starting from the data. As the name of the algorithm suggests, this amounts to minimizing the outflow, determined by the model, of probability from the data. Since the latter is expressed only through the transition probabilities, performing the Monte Carlo is not actually needed, and the gradients of the objective function can be explicitly computed.

For completeness, we briefly summarize the main points in the MPF approach. The model probability density $p$, dependent on a set of parameters $\theta$, it is assumed to evolve according to the master equation:

\begin{equation}
\dot{\mathbf{p}}=\mathbf{\Gamma} \, \mathbf{p}
\end{equation}

The transition probabilities $\Gamma$ are assumed to satisfy the detailed balance condition:

\begin{equation}
\Gamma_{\alpha \rightarrow \beta} \, p_\alpha^{(\infty)}(\theta) = \Gamma_{\beta \rightarrow \alpha} \, p_\beta^{(\infty)}(\theta)
\end{equation}

The Maximum Likelihood estimate $\hat{\theta}$ of the parameters, given the data, is:
\begin{equation}
\hat{\theta}_{\mathrm{ML}} = \underset{\theta}{\mathrm{argmin}} \, D_{\mathrm{KL}}(\mathbf{P}^{(0)} || \mathbf{P}^{(\infty)}(\theta))
\end{equation}

MPF proposes to consider instead, for infinitesimal $\epsilon$:
\begin{equation}
\hat{\theta}_{\mathrm{MPF}} = \underset{\theta}{\mathrm{argmin}} \, D_{\mathrm{KL}}(\mathbf{p}^{(0)} || \mathbf{p}^{(\epsilon)}(\theta))
\end{equation}

It was proven in \cite{MPF} that:
\begin{equation}
D_{\mathrm{KL}}(\mathbf{p}^{(0)} || \mathbf{p}^{(\epsilon)}(\theta)) \simeq \frac{\epsilon}{N_{\mathrm{Data}}} 
\sum_{_{\tiny{
\begin{aligned}
     \alpha \in  & \mathrm{Data}\\[-.2\baselineskip]
      \beta \notin & \mathrm{Data}
\end{aligned}}}} \Gamma_{\alpha \to \beta}
\end{equation}

The MPF strategy is therefore to search for:
\begin{equation}
\min_\theta \sum_{\tiny 
{\begin{aligned}
     \alpha \in  & \mathrm{Data}\\[-.2\baselineskip]
      \beta \notin & \mathrm{Data}
\end{aligned}}
}
\Gamma_{\alpha \rightarrow \beta}
\end{equation}

\subsection{Spiking network and Pseudo-Boltzmann learning}
\label{sec:spiking}
The single neuron obeys the sub-threshold dynamics:
\begin{eqnarray}
\label{IF}
\dot{V_i} &=& \frac{-V_i}{\tau}+\sum_j J_{ij} \sum_{k_j} \delta(t-t_{k_j}-d_{i j})-g_{SFA} c_i + I_i \nonumber \\
\dot{c_i} &=& \frac{-c_i}{\tau_{SFA}}+\sum_{k_i} \delta(t-t_{k_i}) ,
\end{eqnarray}
\noindent with the additional condition that if $V_i \geq V^{th}$, the emission of a spike is recorded and $V_i{}$ is reset to a value $H$ for the duration of a refractory period $t^{ref}$. In Eq.~\ref{IF}, $V_i$ is the membrane potential of neuron $i$, $\tau$ is the membrane time constant, $J_{ij}$ is the synaptic efficacy from neuron $j$ to neuron $i$, $k_j$ label spikes emitted by neuron $j$, $d_{ij}$ is the spike transmission delay from $j$ to $i$. The neuron model includes spike-frequency adaptation (SFA) in the form of a calcium-dependent inhibitory (potassium) current: $g_{SFA}$ is the strength of SFA, $c_i$ represents the intra-cellular calcium concentration, and $\tau_{SFA}$ is a characteristic time of calcium dynamics. The $c_i{}$ variable acts as an integrator of the spiking activity of the neuron $i{}$, such that SFA implements an activity-dependent self-inhibition. $I_i$ is an external current applied to the neuron, implemented as a Poisson process with rate $\nu_i^{ext}$.

The network comprises 64 modules. Each module includes $n_{E}=32$ excitatory ($E$) neurons and $n_{I}=16$ inhibitory ($I$) neurons. $E$/$I$ neurons within a module are connected with probabilities $c_{XY}$, $X,Y=$ $E$/$I$; each $E$/$I$ neuron also receives $C_{E/I}^{ext}$ external spike trains through synapses with mean $J_{E/I}^{ext}$, and with rate $\nu^{ext}_{init}$ ({\it init} refers to the fact that external rates are set to an initial value, then they are optimized during the learning process - see below). Delays $d_{ij}$ are drawn from an exponential distribution between $d_{XY}^{min}$ $d_{XY}^{max}$, $X,Y=$ $E$/$I$, where $d_{XY}^{min}=0.1$ms always.  The values of these parameters are given in Table~\ref{table.parameters}. 
Inter-module synapses are only between excitatory neurons, and can be positive or negative, which can can be viewed as an effective way to also incorporate the excitatory input to inhibitory neurons; their mean efficacies are determined through the learning procedure described below.

All the synapses in the network, given their (learnt or assigned) mean efficacies, are drawn from a Gaussian distribution $25 \%$ relative variance.

The probability of connection between two excitatory neurons belonging to different modules is $0.5$ while the delays for inter-module communication have $d_{EE}^{max}=50$ms. 

To choose the intra-module synaptic connectivity such that the isolated module is approximately bistable, we used predictions from mean-field theory \cite{MattiaEtAlJNeurosci2013}. The bistable behavior and a narrow range of firing rates for the low and high activity states proved to be robust against quenched noise due to different realizations of the synaptic matrices. This allowed us to define a threshold $\theta_{\mathrm{bin}}$ on the firing activity to binarize the dynamic states of the modules (the binarization is actually performed on a smoothed version of activity by averaging over a 20 ms window). In this way, the dynamics of the network is represented by a sequence of binary vectors corresponding to the high 
or low state of all modules. The procedure to construct the inter-module connectivity (see below) preserves to a large extent the modules' bistability for the interconnected network. 

Numerical simulations were performed using the event-driven simulator we described in \cite{mattia2000efficient}.

\begin{table}[!ht]
\caption{\bf{Parameters of the neural modules}} 
\centering 
\begin{tabular}{|c|c|}
\hline
Parameter & Value \\ [0.5ex] 
\hline 
$n_E$ / $n_I{}$ & $32$ / $16$ \\ \hline
$\tau_E$ / $\tau_I$ & $20$ / $10$ ms \\ \hline
$t^{ref}_E$ / $t^{ref}_I$ & $2$ / $1$ ms \\ \hline
$H$ / $V^{th}$ & $15$ / $20$ mV \\ \hline
$c_{EE}$ / $c_{IE}$ / $c_{EI}$ / $c_{II}$ & $0.5$ / $0.5$ / $1$ / $1$ \\ \hline
$J_{EE}$ / $J_{IE}$ / $J_{EI}$ / $J_{II}$ & $0.72$ / $1$ / $-2$ / $-0.0012$ mV \\ \hline
$d^{max}_{EE}$ / $d^{max}_{IE}$ / $d^{max}_{EI}$ / $d^{max}_{II}$ & $21$ / $21$ / $1$ / $1$ ms \\ \hline
$\nu^{ext}_{init}$ & $12$ kHz \\ \hline
$C^{ext}_E$ / $C^{ext}_I$  & $600$ / $400$  \\ \hline
$J^{ext}_E$ / $J^{ext}_I$  & $0.320$ /$0.111$ mV \\ \hline
$\tau_{SFA}$ & $[62.5 - 4000]$ ms \\ \hline
$g_{SFA}$ & $25/\tau_{SFA}$ \\ \hline
$\theta_{\mathrm{bin}}$ & 40 Hz \\ \hline
\end{tabular}
\label{table.parameters} 
\end{table}

The procedure to construct inter-module synaptic connectivity aims to ``store'' prescribed patterns of activities using correlations between the binarized network state vectors to build an effective Hopfield-like synaptic matrix. To store patterns in the modular network we start from a Hopfield model with given $P$ stored patterns, from the simulation of which we estimate the mean single spin magnetization $m_p^{\mathrm{Hopf}}\equiv \langle \sigma_p \rangle_t$, and spatial spin-spin correlation functions $c_{pq}^{\mathrm{Hopf}}\equiv\langle \sigma_p \sigma_q \rangle_t-\langle  \sigma_p \rangle_t \langle \sigma_q \rangle_t$, where $\langle\rangle_t$ is the average over the whole time series. 

We then iteratively minimize the difference between $m_p^{\mathrm{Hopf}}$ and $c_{pq}^{\mathrm{Hopf}}$, and the corresponding quantities $m_p$ and $c_{pq}$ measured from the sequence of binarized states of the multi-modular spiking network (such that $p$ and $q$ are module labels). Such minimization is performed adopting an educated guess inspired by Boltzmann learning that would be appropriate for a binary spin system, and that we verify \textit{a posteriori}. Specifically we implement the following procedure:
\begin{itemize}
\item Run a long simulation of the multi-modular system
\item Binarize the activity of each module, obtaining a time-series of binary state vectors, from which
\item Measure the spatial connected correlations $c_{pq}$ and magnetizations $m_p$
\item Perform a step of the pseudo-Boltzmann iteration over the values of $J_{pq}$ and $\nu_p^{ext}$, where $J_{pq}$ is the average synaptic efficacy from neurons belonging to module $q$ to neurons belonging to module $p$; and $\nu_p^{ext}$ is the common value of $\nu^{ext}$ for all neurons in module $p$
\begin{eqnarray}
\Delta J_{pq}^{(t)} & = & -\mathrm{sign}(c_{pq}^{(t)} - c_{pq}^{\mathrm{Data}}) \, \Delta_{pq}^{(t)} \nonumber \\ 
\Delta \nu^{(t)}_{{\mathrm{ext}}_{p}} & = & -\mathrm{sign}(m_{p}^{(t)} - m_{p}^{\mathrm{Data}}) \, \Delta_{p}^{(t)}  
\end{eqnarray}

with the learning rates $\Delta_{pq}^{(t)}$ and $\Delta \nu^{(t)}$ determined by the Rprop algorithm:

\begin{equation}
\Delta_{pq}^{(t)}=
\begin{cases}
\mathrm{min}(\eta^{+} \Delta_{pq}^{(t-1)},\Delta_{\mathrm{max}}) \,\,\, & \mathrm{if} \, (c_{pq}^{(t-1)} - c_{pq}^{\mathrm{Data}})(c_{pq}^{(t)} - c_{pq}^{\mathrm{Data}}) >0    \\
\mathrm{max}(\eta^{-} \Delta_{pq}^{(t-1)},\Delta_{\mathrm{min}}) \,\,\, & \mathrm{if} \, (c_{pq}^{(t-1)} - c_{pq}^{\mathrm{Data}})(c_{pq}^{(t)} - c_{pq}^{\mathrm{Data}})<0    \\
\Delta_{pq}^{(t-1)} \,\,\, & \mathrm{otherwise} \\
& 0 \, < \, \eta^- \, < \, 1 \, < \, \eta^+\\
\end{cases} 
\label{eqRprop1}
\end{equation}

\begin{equation}
\Delta_{p}^{(t)}= \begin{cases}
\mathrm{min}(\eta^{+} \Delta_{p}^{(t-1)},\Delta_{\mathrm{max}}) \,\,\, & \mathrm{if} \, (m_{p}^{(t-1)} - m_{p}^{\mathrm{Data}})(m_{p}^{(t)} - m_{p}^{\mathrm{Data}}) >0    \\
\mathrm{max}(\eta^{-} \Delta_{p}^{(t-1)},\Delta_{\mathrm{min}}) \,\,\, & \mathrm{if} \, (m_{p}^{(t-1)} - m_{p}^{\mathrm{Data}})(m_{p}^{(t)} -m_{p}^{\mathrm{Data}})<0    \\
\Delta_{p}^{(t-1)} \,\,\, & \mathrm{otherwise} \\
& 0 \, < \, \eta^- \, < \, 1 \, < \, \eta^+\\
\end{cases} 
\label{eqRprop2}
\end{equation}

In essence, Rprop adapts the learning rate for each parameter only using the sign of the derivatives: when the sign of the derivative does not change between two successive iteration steps, learning should speed up, while when the sign of the derivative changes it should slow down.

\end{itemize}

We checked that, although every simulation during the training procedure starts from the same exact condition, discarding the first $5 \times \tau_{SFA}$ steps of each run is enough to ensure that no appreciable correlations are detectable between different realizations.

\subsection{Complexity measure of dynamics in the centroid space}
\label{sec:lz}

After performing MS clustering on the time series generated by a simulation of the multi-modular spiking network with given $\tau_{SFA}$ and $g_{SFA}$, to each of the 64-dimensional vectors containing the modules' firing activities we substitute the label of the centroid that vector belongs to. In this way, the multi-dimensional dynamics is converted into a symbolic sequence of centroid labels. 
 
From such label sequence we measure the matrix of transition probabilities between all pairs of centroids, and generate surrogate label sequences as Markov processes with the same transition probabilities as the actual sequence. While, by construction, the surrogate sequences are memoryless, this may not be the case for the actual one. We focus on such possible memory effects, which are in general interesting to uncover, and for this purpose we adopt a measure typically used to evaluate the complexity of symbolic sequences, the Lempel-Ziv (LZ) complexity, which is a measure of compressibility, suitably normalized in order to eliminate trivial dependence on the length of the sequence.

The LZ complexity is computed as \cite{lempel1976complexity} 
\begin{equation}
{\cal C}_\mathrm{LZ} = \frac{S_\mathrm{LZ}}{|S| \, \log(|A|)/\log(|S|)}
\end{equation}
where $S_\mathrm{LZ}$ is the length of the compressed sequence, $|S|$ is the length of the sequence, $|A|$ is the length of the alphabet.

And the relative complexity index is defined as:

\begin{equation}
\mathrm{R} = \frac{\cal{C}_\mathrm{LZ}^\mathrm{Markov} - \cal{C}_\mathrm{LZ}^\mathrm{Sample}}{\cal{C}_\mathrm{LZ}^\mathrm{Markov}}
\end{equation}

where $\cal{C}_\mathrm{LZ}^\mathrm{Sample}$ is the LZ complexity of the actual centroids sequence, and $\cal{C}_\mathrm{LZ}^\mathrm{Markov}$ is the average LZ complexity of the surrogate Markov centroids sequences.

\section{Acknowledgements}
We thank Maurizio Mattia and Cristiano Capone for a critical reading of a previous version of the manuscript, and Giorgio Parisi for useful discussions in the early stage of the work.
We are grateful to Paolo Barucca for several constructive interactions during the initial phase of the work.
Gabriel Baglietto acknowledges financial support from CONICET.
Paolo Del Giudice was supported in part by the EU Grant WaveScalES (EC FET Flagship HBP SGA1 720270)

\end{document}